\documentclass[reprint,aps,pre]{revtex4-2}

\usepackage{graphicx}
\usepackage{amsmath,amssymb,bm}
\usepackage{hyperref}
\usepackage{color}

\begin{document}

\title{Sum-of-squares bounds on correlation functions in a minimal model of turbulence}
\date\today

\author{Vladimir Parfenyev}\email{parfenius@gmail.com}
\affiliation{Landau Institute for Theoretical Physics, 142432 Chernogolovka, Russia}
\affiliation{National Research University Higher School of Economics, Faculty of Physics, 101000 Moscow, Russia}

\author{Evgeny Mogilevskiy}\email{e.i.mogilevskiy@gmail.com}
\affiliation{Lomonosov Moscow State University, Faculty of Mechanics and Mathematics, 11992 Moscow, Russia}
\affiliation{Weizmann Institute of Science, 76100 Rehovot, Israel}

\author{Gregory Falkovich}\email{gregory.falkovich@weizmann.ac.il}
\affiliation{Weizmann Institute of Science, 76100 Rehovot, Israel}
\affiliation{Landau Institute for Theoretical Physics, 142432 Chernogolovka, Russia}


\begin{abstract}
We suggest a new computer-assisted approach to the development of turbulence theory. It allows one to impose lower and upper bounds on correlation functions using sum-of-squares polynomials. We demonstrate it on the minimal cascade model of two resonantly interacting modes, when one is pumped and the other dissipates. We show how to present correlation functions of interest as part of a sum-of-squares polynomial using the stationarity of the statistics. That allows us to find how the moments of the mode amplitudes depend on the degree of non-equilibrium (analog of the Reynolds number), which reveals some properties of marginal statistical distributions. By combining scaling dependence with the results of direct numerical simulations, we obtain the probability densities of both modes in a highly intermittent inverse cascade. We also show that the relative phase between modes tends to $\pi/2$ and $-\pi/2$ in the direct and inverse cascades as the Reynolds number tends to infinity, and derive bounds on the phase variance. Our approach combines computer-aided analytical proofs with a numerical algorithm applied to high-degree polynomials.
\end{abstract}

\maketitle

\section{Introduction}

Many systems in nature receive and dissipate energy on very different scales, having conservative dynamics in between, and the energy is transferred across the scales through a turbulent cascade. Among the best-known examples are isotropic fluid turbulence~\cite{frisch1995turbulence} and surface waves in the ocean~\cite{zakharov2012kolmogorov}. The complexity of such systems makes them difficult to describe in detail, so it makes sense to consider simpler dynamical models  to deepen our understanding of the statistical properties and energy transfer in such highly non-equilibrium systems~\cite{biferale2003shell,obukhov1969integral,Falkovich2021Fibonacci}.


A minimal model, which still captures the basic properties of turbulent cascades, is a system of two resonantly interacting oscillators whose natural frequencies differ by a factor of two~\cite{vladimirova2021second}. This system allows studying direct and inverse cascades with the energy flux directed towards either higher or lower frequencies. The system has one non-dimensional governing parameter $\chi$, which plays the role of the Reynolds number. In what follows, we are mostly interested in the regime when this parameter is large, and the probability distribution tends to be singular. The analytical studies in this limit resulted in the steady-state probability density for the direct cascade, while constructing the probability density for the inverse cascade turns out to be tricky~\cite{vladimirova2021second}. 



Here we apply a complementary approach to study the mode statistics in this system. We exploit the polynomial nature of dynamic equations (common for practically all turbulent systems), which allows us to impose inequalities on the correlation functions. The main idea is to find a non-negative polynomial expression, $\phi(\bm x) - L + F(\bm x) \geq 0$ that combines the correlation function $\phi(\bm x)$ to be bounded, the constant value of the bound $L$, and an auxiliary polynomial function $F(\bm x)$ that has zero mean value $\langle F(\bm x) \rangle=0$ in a statistically steady state, which entails the inequality $\langle \phi(\bm x) \rangle \geq L$~\cite{chernyshenko2014polynomial,fantuzzi2016bounds}. The essence of the approach is to construct the function $F(\bm x)$ so that the value of the lower bound $L$ is as large as possible. Although testing a polynomial expression $\phi(\bm x) - L + F(\bm x)$ for non-negativity is NP-hard algorithmic task, there exist numerical procedures~\cite{parrilo2000structured, parrilo2003semidefinite, prajna2002introducing, lofberg2004yalmip, legat2017sos} that solve the problem with a more strict requirement on the proposed expression to be sum-of-squares (SoS) of other polynomials. Moreover, these procedures allow one to maximize the value $L$ of the bound using some ansatz for the auxiliary function $F(\bm x)$. If the ansatz is simple enough, the bound can be found analytically, while a computer algorithm can be used in advance to suggest the optimal form of the SoS polynomial. The upper bound $\langle \phi(\bm x) \rangle \leq U$ can be constructed in a similar way. Recent examples of the application of SoS programming to study dynamical systems can be found in Refs.~\cite{chernyshenko2014polynomial, fantuzzi2016bounds, papachristodoulou2002construction, papachristodoulou2005tutorial, tan2006stability, goulart2012global}.


The rest of the paper is organized as follows. In Section~\ref{sec:SoS}, we remind the general theory behind SoS optimization and then apply the method for the two-mode system. In Section~\ref{sec:direct2}, we present the results obtained by SoS programming for the direct cascade and compare them with the analytics and direct numerical simulations (DNS). It turns out that the upper and lower bounds for the moments of pumped and dissipating modes are close to each other and differ by less than a percent in the limit of a large Reynolds number. This allows us to determine not only their scaling dependence but also numerical values with accuracy comparable to DNS. The method can also be used to study correlations between modes, and we show that the relative phase between the modes tends to $\pi/2$ and determine upper and lower bounds for its root-mean-square fluctuations.

After the validation, in Section~\ref{sec:inverse2}, we use the method for the inverse cascade where the probability density is unknown a priori. Analytically, we obtained only lower bounds for the correlation functions, but they demonstrate scaling consistent with the results of DNS. A numerical algorithmic analysis of the high moments of the dissipating mode leads to scaling $\langle n_1^k\rangle \propto \chi^{k-1}$, which is a fingerprint of intermittency, where $n_1$ is the mode intensity and $\chi$ is the Reynolds number. Based on this observation, we were able to shed light on the structure of the distribution function of this mode -- the DNS results fall on the universal curve for different values of $\chi$, which has a form close to a power law with an exponential cutoff. As for the pumped mode, its statistics are close to Gaussian, and we analytically found the lower bound for its intensity $\langle n_2 \rangle \geq 5 \chi/16$ that is close to DNS. We also show that the relative phase between modes tends to $-\pi/2$ in the limit $\chi \gg 1$ and estimate the rate of this transition. Finally, we summarize and discuss our findings in Section~\ref{sec:conclusion}.

\section{Theoretical Framework}\label{sec:SoS}

This section briefly explains how sum-of-squares optimization can impose inequalities on correlation functions in stochastic systems with polynomial dynamics. The presentation follows Ref.~\cite{fantuzzi2016bounds}, where a more detailed discussion can be found.

Let us consider a stochastic dynamical system
\begin{equation}
    \dot{x_i} = f_i (\bm x) + \sigma_{ij}(\bm x) \xi_j (t), \quad \bm x \in \mathbb{R}^n, \; \bm \xi \in \mathbb{R}^m,
\end{equation}
where $f_i (\bm x)$ and $\sigma_{ij}(\bm x)$ are polynomial, $\xi_i(t)$ is a Gaussian noise with zero mean $\langle \xi_i(t) \rangle = 0$ and the variance $\langle \xi_i (t) \xi_j (t') \rangle =\delta_{ij} \delta(t-t')$, and here and below we sum over repeated indices. We assume that the system has reached a statistical steady-state, and then the probability density function $\rho(\bm x)$ satisfies the stationary Fokker-Planck equation
\begin{equation}
\frac{1}{2} \partial_i [\sigma_{ij} \partial_k (\sigma_{kj} \rho)] - \partial_i (f_i \rho) = 0.
\end{equation}
We also assume that the solution to this equation is unknown. We wish to prove a constant lower bound $\langle \phi(\bm x) \rangle \geq L$ for some polynomial correlation function $\phi(\bm x)$, where the angle brackets mean the averaging over the probability density $\rho(\bm x)$.

For this purpose, we consider an auxiliary function $Q(\bm x)$, which does not grow fast when $|\bm x| \to \infty$, so that all the moments considered below are finite. Performing integration by parts and neglecting boundary terms, one can show that $\langle \frac{1}{2} \sigma_{kj} \partial_k (\sigma_{ij} \partial_i Q) + f_i \partial_i Q\rangle = 0$. The idea is to properly design $Q(\bm x)$ so that
\begin{equation}\label{eq:main_ineq}
\left\langle \frac{1}{2} \sigma_{kj} \partial_k (\sigma_{ij} \partial_i Q) + f_i \partial_i Q + \phi - L \right\rangle \geq 0 \ .
\end{equation}
Evaluation of the expectation in expression (\ref{eq:main_ineq}) requires knowing $\rho(\bm x)$, but it is sufficient for the inequality to hold point-wise for all $\bm x$.

Checking a polynomial expression for non-negativity is NP-hard algorithmic task, therefore to reduce computational complexity, it can be replaced with a semidefinite programming (SDP) problem of checking the stronger condition that expression (\ref{eq:main_ineq}) belongs to a set of polynomials that are sum-of-squares (SoS).
Finally, to find the maximum value of the constant $L$, we arrive at the following optimization problem
\begin{equation}\label{eq:V_L}
 \max_{Q(\bm x)} L: \; \frac{1}{2} \sigma_{kj} \partial_k (\sigma_{ij} \partial_i Q) + f_i \partial_i Q + \phi - L \in SoS.
\end{equation}
One specifies an ansatz for the auxiliary function $Q(\bm x)$ with undetermined coefficients and then solves the problem using the software packages such as SOSTOOLS~\cite{prajna2002introducing}, YALMIP~\cite{lofberg2004yalmip} or SumOfSquares.jl~\cite{legat2017sos} with one of the appropriate SDP-solvers~\cite{andersen2009mosek, fujisawa2008sdpa, sturm1999using, tutuncu2003solving}.
The upper bounds $\langle \phi(\bm x) \rangle \leq U$ can be obtained in a similar way by considering the optimization problem
\begin{equation}\label{eq:V_U}
 \min_{Q(\bm x)} U: \; - \frac{1}{2} \sigma_{kj} \partial_k (\sigma_{ij} \partial_i Q) - f_i \partial_i Q - \phi + U  \in SoS.
\end{equation}
If the ansatz is simple enough, the bounds $L$ and $U$ can be found analytically, and the computer algorithm suggests the optimal form of the SoS polynomials. The more complex ansatz leads to more rigorous bounds; however, complex ansatz requires more computational efforts, and the numerical algorithm could fail to find an optimal solution.

\section{Direct cascade}\label{sec:direct2}

\begin{figure*}
\includegraphics[width=\textwidth]{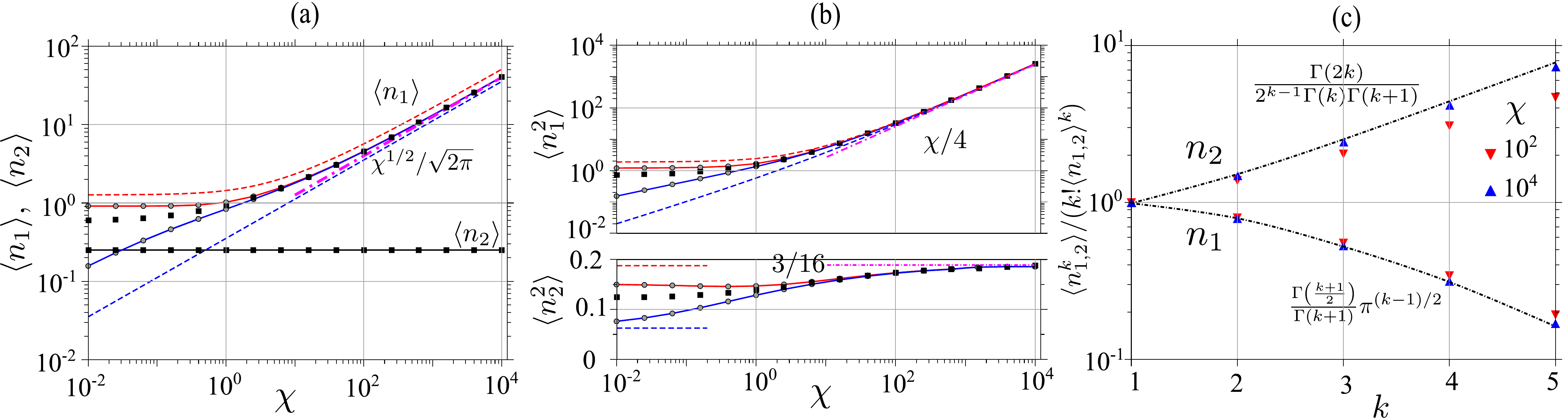}
\caption{\label{fig:direct_n} Average values of mode intensities (a), their second (b), and higher moments (c) for the direct cascade. Solid and dashed lines show numerical ($d=10$) and analytical ($d=4$) results for the upper (red) and lower (blue) bounds. The square markers present results of DNS, and dash-dotted lines correspond to asymptotics following from relation (\ref{eq:direct_large_chi}). Expressions for the asymptotic dependence are shown in the panels.}
\end{figure*}

The direct cascade for the two-mode system is governed by the following non-dimensional system of equations~\cite{vladimirova2021second}
\begin{eqnarray}
&\displaystyle \dot{b}_1=-\frac{2ib_1^*b_2}{\sqrt{\chi}} +\xi(t),&\label{eq:direct_b1}\\
&\displaystyle \dot{b}_2=-\frac{ib_1^2}{{\sqrt{\chi}}} -b_2,\label{eq:direct_b2}&
\end{eqnarray}
where $b_1$ and $b_2$ are complex envelopes of low- and high-frequency modes, respectively, $\xi(t)$ is a complex Gaussian random force with zero mean $\langle \xi(t) \rangle = 0$ and the variance $\langle \xi(t) \xi^*(t') \rangle = \delta(t-t')$, and the asterisk means complex conjugate. In other words, real and imaginary parts of $\xi(t)$ are independent white noises having zero mean values and intensities $1/2$. The parameter $\chi$ is the only control parameter in the system, and it quantifies the dissipation relative to interaction strength. In the statistical steady-state, the energy input rate equals the energy flux from the first mode to the second and the dissipation rate. For the amplitude of the dissipating mode and the flux, one finds $\langle n_2 \rangle \equiv \langle |b_2|^2 \rangle = 1/4$, $\langle J \rangle \equiv \langle i b_1^2 b_2^* + c.c. \rangle/ \sqrt{\chi} = -1/2$.

In the limit $\chi \ll 1$, the interaction between modes is strong, and the energy transfer is fast, so one may expect that the occupation numbers are close to energy equipartition $\langle|b_1|^2\rangle\equiv\langle n_1 \rangle = 2 \langle n_2 \rangle$ corresponding to thermal equilibrium~\cite{vladimirova2021second}. The DNS shows that the marginal distributions of the mode amplitudes are indeed close to the Gaussian statistics with the corresponding occupation numbers. Still, the phase between the modes $\theta = \arg(b_1^2 b_2^*)$ is unevenly distributed, which indicates the presence of correlations between the modes~\cite{vladimirova2021second}.

In the opposite limit of weak interaction and strong noise, $\chi \gg 1$, one expects that the driven mode needs much higher amplitude to provide for the flux: $\langle n_1 \rangle \gg \langle n_2 \rangle$. In other words, the mode statistics are far from thermal equipartition. In this case, the asymptotic analytical solution for the whole distribution function was argued to be singular~\cite{vladimirova2021second}:
\begin{equation}\label{eq:direct_large_chi}
\mathcal{P}(b_1,b_2)=\frac{2^{3/2}}{\pi^{3/2} \chi^{1/2}}\exp\left(
-\frac{2}{\chi} |b_1|^4
\right)
\delta\left(
b_2+ \frac{i b_1^2}{\sqrt{\chi}}
\right)\ .
\end{equation}
We will use this result later for comparison with  the results of our approach, which provides mutual validation.

Let us now apply the SoS optimization method described in the previous section. To proceed, we need to specify an ansatz for the auxiliary function $Q$. We have found that the numerical procedure works better if $Q$ is a polynomial with respect to $n_1,\ n_2,\ J$ and takes the monomials $n_1^i n_2^j J^k$ with the total degree of mode amplitudes less than $d \geq 2i+2j+3k$. A more general ansatz does not improve estimates for the correlation functions, but the algorithm is less stable due to the expansion of the optimization space. Let us emphasize that for small values of $d$, the bounds for the correlation functions can be obtained analytically, the computer algorithm operates for higher values of $d$ and improves the result.

We begin to present our results with the intensity of the pumped mode. For $d=4$ we obtain analytically~\cite{SM}
\begin{equation}\label{eq:direct_n1}
    \frac{\chi^{1/2}}{2 \sqrt{2}} \leq \langle n_1 \rangle \leq \dfrac{1}{2} \left(3 -\sqrt{3} + \sqrt{12 - 6 \sqrt{3}+\chi} \right),
\end{equation}
and in the limit $\chi \gg 1$, this implies ${\chi^{1/2}}/{2 \sqrt{2}} \leq\langle n_1 \rangle \leq \sqrt{\chi}/2$.
Inequality (\ref{eq:direct_n1}) means, in particular, that the intensity of the pumped mode is much greater than the intensity of the dissipated mode, and $\langle n_1 \rangle/\langle n_2 \rangle \propto \sqrt{\chi}$.
As the parameter $d$ increases, the numerically found upper and lower bounds for $d=10$ approach very closely the asymptotic $\langle n_1 \rangle = \sqrt{\chi/2 \pi}$ following from expression (\ref{eq:direct_large_chi}), see Fig.~\ref{fig:direct_n}a.
In the opposite case $\chi \ll 1$, the upper bound $\langle n_1 \rangle \leq 3 - \sqrt{3}$ does not depend on $\chi$ in qualitative agreement with DNS. The lower bound is not tight, and increasing the parameter $d$ does not qualitatively change its behavior.

Similar results are also obtained for higher moments of $n_1$. For the second moment and $d=4$ we analytically find~\cite{SM}
\begin{equation}
    \chi^2 r_l(\chi) \leq \langle n_1^2 \rangle \leq \frac{\chi}{4} + \chi^2 r_u (\chi),
\end{equation}
where $r_l (\chi)$ is the largest real root of the equation $1 - \chi^2 r + 8 \chi^3 r^2 - 16 \chi^4 r^3 = 0$ and $r_u (\chi)$ is the smallest positive real root of the equation $1 + 16 \chi + (4 \chi + 152 \chi^2) r + (348 \chi^3 - 32 \chi^4) r^2 - 216 \chi^5 r^3 + 16 \chi^7 r^4 = 0$. In the limit $\chi \gg 1$, the upper and lower bounds coincide with each other and therefore $\langle n_1^2 \rangle \to \chi/4$, while in the opposite case $\chi \ll 1$, one obtains $2^{-4/3} \chi^{2/3} \leq \langle n_1^2 \rangle \lesssim 1.88$, where the upper bound reflects scaling consistent with DNS, see Fig.~\ref{fig:direct_n}b. For even higher moments of $n_1$, the values in the limit $\chi \gg 1$ can be determined numerically with good accuracy since the upper and lower bounds tend to each other as $d$ increases. Analyzing the values of the moments, one can conclude that the statistics of the mode $b_1$ are essentially non-Gaussian in agreement with expression (\ref{eq:direct_large_chi}), see Fig.~\ref{fig:direct_n}c. This kind of analysis can help one to guess the marginal distribution function if it is not known a priori.

Next, we turn to the dissipating mode. Its mean intensity is determined exactly from the energy balance condition, $\langle n_2 \rangle = 1/4$, and the upper and lower bounds for $\langle n_2^2 \rangle$ are shown in Fig.~\ref{fig:direct_n}b. In the limit $\chi \gg 1$, the reasonable bounds are obtained for relatively large values of $d \geq 6$, so we do not present analytical results. In the opposite case $\chi \ll 1$, the bounds demonstrate scaling consistent with DNS, but they are not close to each other. Analytically we obtain $1/16 \leq \langle n_2^2 \rangle \leq 3/16$ for $d=4$ and $\chi \ll 1$~\cite{SM}. The analysis of higher moments for $\chi \gg 1$ is presented in Fig.~\ref{fig:direct_n}c, and the results are in agreement with the marginal distribution $P(b_2) = \frac{2^{1/2}}{\pi^{3/2}|b_2|} e^{-2|b_2|^2}$ following from expression (\ref{eq:direct_large_chi}).

\begin{figure}
\includegraphics[width=0.75\linewidth]{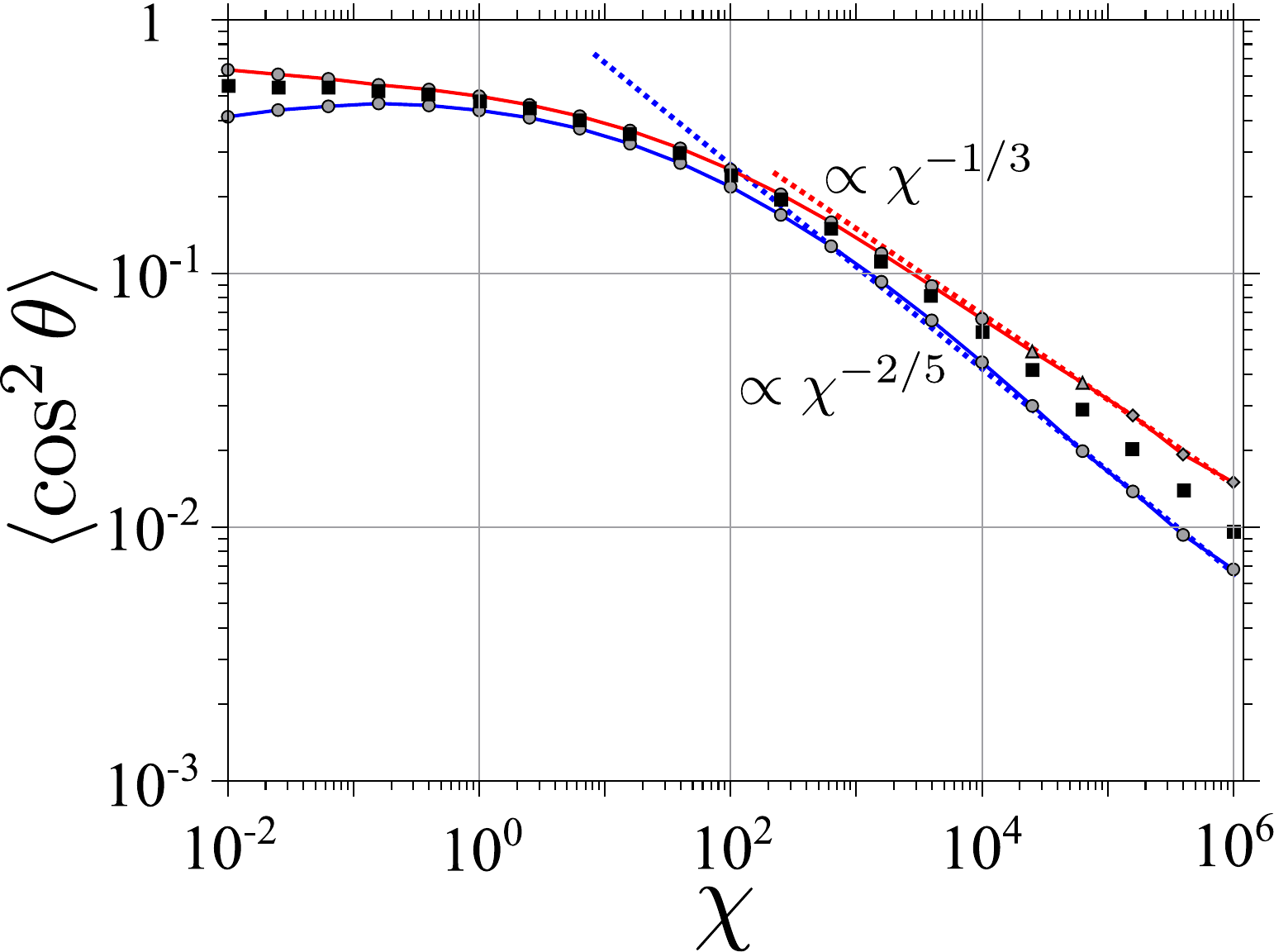}
\caption{\label{fig:direct_cos2} Relative phase between modes for the direct cascade: numerically obtained upper (red) and lower (blue) bounds (circles, triangles, and diamonds are for $d=10,\ 12,\ 14)$; the square markers present results of DNS.}
\end{figure}

The SoS programming method can also be used to analyze correlations between modes. To estimate the relative phase between modes, we rewrite the initial equations (\ref{eq:direct_b1})-(\ref{eq:direct_b2}) in terms of real variables $\rho_1=|b_1|$, $\rho_2=|b_2|$, $\theta=\arg(b_1^2b_2^*)$:
\begin{eqnarray}
&\dot{\rho_1}=-\dfrac{2\rho_1\rho_2\sin\theta}{\sqrt{\chi}} +\dfrac{1}{4\rho_1}+\dfrac{\zeta_1(t)}{\sqrt{2}},&\\    &\dot{\rho_2}=\dfrac{\rho_1^2 \sin\theta}{\sqrt{\chi}} - \rho_2,&\\
&\dot{\theta}=\dfrac{\rho_1^2-4\rho_2^2}{\rho_2 \sqrt{\chi}}\cos\theta+\dfrac{\sqrt{2}\zeta_2(t)}{\rho_1},
\end{eqnarray}
where the overall phase drops out, and $\zeta_{i}(t)$ is a real Gaussian noise with zero mean and the pair correlation function $\langle \zeta_i(t) \zeta_j(t') \rangle = \delta_{ij}\delta(t-t')$~\cite{SM}. Despite the right-hand sides of these equations are not polynomial, for any functions $Q$ and $\phi$ that are polynomial with respect to $\rho_1$, $\rho_2$, $\sin\theta$ the expressions in left-hand sides of Eqs.~(\ref{eq:V_L}), (\ref{eq:V_U}) are polynomials divided by $\rho_1^2\rho_2>0$. Thus, to apply the algorithm, it is sufficient to multiply the expression in the optimization problems (\ref{eq:V_L}), (\ref{eq:V_U}) by $\rho_1^2\rho_2>0$ to return it into the class of polynomial functions, see details in Ref.~\cite{SM}. Now the ansatz for the function $Q$ is a polynomial of $\rho_1,\ \rho_2,\ \sin\theta$ of the power of $d$. We have also found that the numerical procedure gives better results if we solve the optimization problem with the additional constraint $1 - \sin^2 \theta \geq 0$~\cite{SM}.


\begin{figure*}
\includegraphics[width=\textwidth]{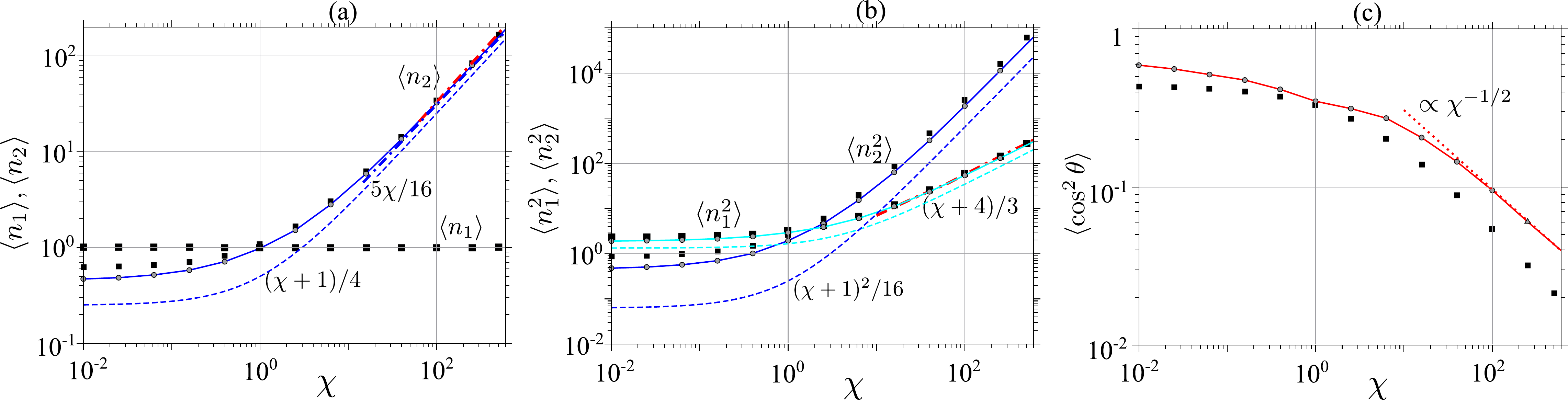}
\caption{\label{fig:inverse_n} Average values of mode intensities (a), their second moments (b), and the relative phase (c) for the inverse cascade. Solid and dashed lines show numerical ($d=10$) and analytical results for upper and lower bounds. The square markers present results of DNS, and dash-dotted lines correspond to asymptotic bounds in the limit $\chi \gg 1$. Expressions for the analytic results are shown in the panels.}
\end{figure*}

Fig.~\ref{fig:direct_cos2} shows the upper and lower bounds on $\langle \cos^2 \theta \rangle$. The value of $\langle \cos^2 \theta \rangle \to 0$ as $\chi \to \infty$, which means that the phase $\theta \to \pi/2$ (the point $\theta = -\pi/2$ is not suitable because $\langle J \rangle \equiv \langle -2\rho_1^2\rho_2\sin\theta\rangle/\sqrt{\chi}= -1/2 < 0$). The power-law fit $\langle \cos^2 \theta \rangle \propto \chi^{-q}$ results in $q=1/3$ and $q = 2/5$ for the upper and lower bounds, respectively. We could not determine the bounds analytically since the algorithm gives reasonable estimates only for large values of $d$. The overall picture is that the distribution function of the phase between the modes has a peak at $\theta=\pi/2$, and its width decreases in a power-law manner with the parameter $\chi$. These results complement Ref.~\cite{vladimirova2021second}, in which the narrowing of the phase distribution was not quantified.

\section{Inverse cascade}
\label{sec:inverse2}

Now we turn to the inverse cascade where the high-frequency mode is pumped, and the low-frequency mode dissipates
\begin{eqnarray}
&\displaystyle \dot{b}_1=-\frac{2ib_1^*b_2}{\sqrt{\chi}} -b_1,&\label{eq:inverse_b1}\\
&\displaystyle \dot{b}_2=-\frac{ib_1^2}{{\sqrt{\chi}}} +\xi(t).\label{eq:inverse_b2}&
\end{eqnarray}
In the statistical steady-state, from the energy balance consideration, we find $\langle n_1 \rangle \equiv \langle |b_1|^2 \rangle = 1$ and $\langle J \rangle \equiv \langle i b_1^2 b_2^* + c.c. \rangle/\sqrt{\chi} = 1$, which is true for any value of $\chi$. The energy flux $\langle J \rangle$ is positive that corresponds to the transfer of energy down the frequencies.
As in the case of a direct cascade, in the limit $\chi \ll 1$, one can expect the occupation numbers to be close to the energy equipartition $\langle n_1 \rangle = 2 \langle n_2 \rangle$, although the statistics is not expected to be close to the product of two Gaussians corresponding to thermal equilibrium. In the opposite case $\chi \gg 1$, the system is far from thermal equilibrium, and the pumped mode is expected to have larger intensity, $\langle n_2 \rangle \gg \langle n_1 \rangle$. All these expectations were confirmed by DNS~\cite{vladimirova2021second}.

We found that we can get better bounds for correlation functions if we rewrite dynamic equations (\ref{eq:inverse_b1})-(\ref{eq:inverse_b2}) in terms of real variables $\rho_1=|b_1|$, $\rho_2=|b_2|$, $\theta=\arg(b_1^2b_2^*)$:
\begin{eqnarray}
\label{eq:inv1}
&\dot{\rho_1}=-\dfrac{2\rho_1\rho_2\sin\theta}{\sqrt{\chi}} - \rho_1,&\\
&\dot{\rho_2}=\dfrac{\rho_1^2 \sin\theta}{\sqrt{\chi}} + \dfrac{1}{4 \rho_2} + \dfrac{\zeta_1(t)}{\sqrt{2}},&\\
\label{eq:inv3}
&\dot{\theta}=\dfrac{\rho_1^2-4\rho_2^2}{\rho_2 \sqrt{\chi}}\cos\theta+\dfrac{\zeta_2(t)}{\sqrt{2} \rho_2},
\end{eqnarray}
where the overall phase drops out~\cite{SM}. We also found that in contrast to the direct cascade, it is useful to extend the ansatz for the function $Q$ by adding the term $\log\rho_1$. The term $\log \rho_2$ does not have a noticeable effect on the results. The left-hand sides of Eqs.~(\ref{eq:V_L}), (\ref{eq:V_U}) should be multiplied by $\rho_2^2>0$ so that they become polynomial and we can apply the SDP algorithm to solve the optimization problem~\cite{SM}.


In contrast to the direct cascade, we were able to obtain only lower bounds for the correlation functions in the case of the inverse cascade. The algorithm does not find a feasible solution for the upper bounds, even for a relatively large parameter value $d=10$. Fortunately, the lower bounds are quite informative precisely in the   turbulent limit of large $\chi$ which we focus on. In particular, for the intensity of the pumped mode, we analytically find~\cite{SM}
\begin{equation}\label{eq:inverse_n2}
    \langle n_2 \rangle \geq \dfrac{\chi+1}{4},
\end{equation}
and both asymptotics at $\chi \ll 1$ and $\chi \gg 1$ give a scaling that agrees qualitatively with DNS, see Fig.~\ref{fig:inverse_n}a. Physically, inequality (\ref{eq:inverse_n2}) means that in the limit $\chi \gg 1$, the intensity of the pumped mode is much greater than the intensity of the dissipating mode, $\langle n_2 \rangle/\langle n_1 \rangle \geq \chi/4$. We thus have shown that deviation from the equipartition is much stronger in the inverse cascade than in the direct one (where the ratio is $\propto\sqrt\chi$).

\begin{figure*}
\includegraphics[width=0.95\textwidth]{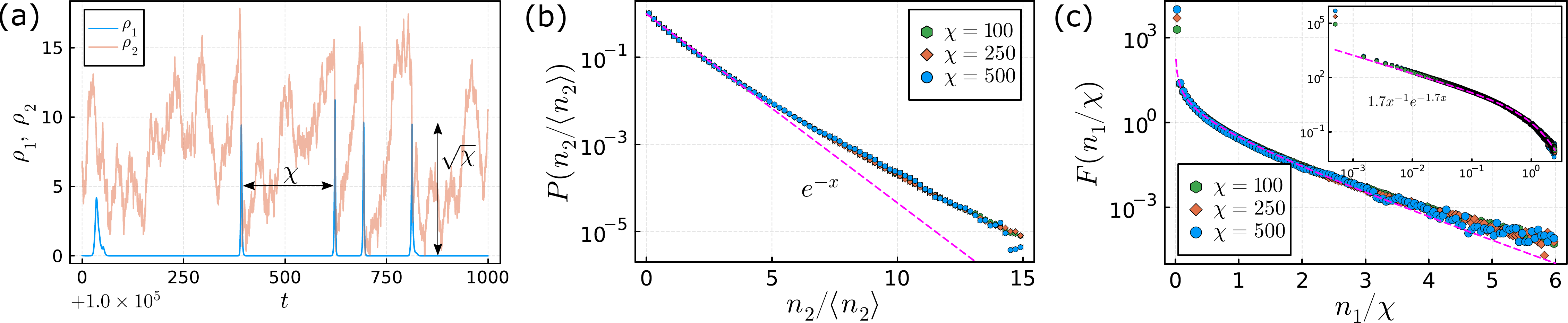}
\caption{\label{fig:4}(a) DNS fragment of the mode dynamics for the inverse cascade demonstrate intermittency of the dissipated mode, $\log_{10} \chi = 2.4$; (b) Probability density for the pumped mode in the inverse cascade. The dashed line corresponds to the exponential distribution, $P(x)=e^{-x}$; (c) Reduced probability density $F(n_1/\chi)$ for the dissipating mode in the inverse cascade, see Eq.~(\ref{eq:p1}). The inset shows the same data in the log-log scale, and the dashed line corresponds to the numerical fit $F(x) = 1.7 x^{-1} e^{-1.7x}$. The bin size is $0.05$ for the main panel and $0.001$ for the inset.}
\end{figure*}

Similarly, we find for the amplitude fourth moment~\cite{SM}
\begin{equation}
\langle n_2^2 \rangle \geq \dfrac{(\chi+1)^2}{16}, \quad \langle n_1^2 \rangle \geq \dfrac{\chi+4}{3}.
\end{equation}
The first condition is trivial and follows from the positive variance of the pumped mode intensity. The second condition means that in the limit $\chi \gg 1$, the ratio $\langle n_1^2 \rangle/\langle n_1 \rangle^2 \geq \chi/3 $, i.e., the statistics of the dissipated mode is intermittent. This agrees with the analysis carried out in Ref.~\cite{vladimirova2021second}, where it was shown that the $\rho_1$ dynamics is a sequence of burst events, see also Fig.~\ref{fig:4}a. Between bursts, the amplitude $\rho_1$ is close to zero, and during short bursts with a duration of order unity,  $\rho_1$ reaches large values $\sim \sqrt{\chi}$. The time interval between bursts is $\sim \chi$, and the correlation functions of $\rho_1$ saturate on bursts. In Fig.~\ref{fig:inverse_n}b, we compare the obtained inequalities with DNS. All asymptotics demonstrate correct scaling with the parameter $\chi$. Note that analytical inequalities are improved by the numerical algorithm when we increase the parameter $d$.

Fig.~\ref{fig:inverse_n}c shows the upper bound for $\langle \cos^2 \theta \rangle \leq U$,
which is equivalent to the lower bound for $\langle \sin^2 \theta \rangle \geq 1-U$. In the limit $\chi \gg 1$, the value of $\langle \cos^2 \theta \rangle \to 0$ and it means that the the phase $\theta \to -\pi/2$, since the value of $\langle J \rangle = \langle -2 \rho_1^2 \rho_2 \sin \theta \rangle/\sqrt{\chi} = 1$. The power-law fit $\langle \cos^2 \theta \rangle \propto \chi^{-q}$ results in $q=1/2$, although the interval is short and with increasing $\chi$ we have to increase $d$ so that the algorithm finds a feasible solution. Compared to the direct cascade, the phase fluctuations in the inverse cascade are smaller for the same value of $\chi$. Based on this observation, we can simplify equations (\ref{eq:inv1})-(\ref{eq:inv3}) by assuming that the relative phase is locked on $\theta=-\pi/2$. Then, we obtain
\begin{eqnarray}
\label{eq:inv_simple_1}
&\dot{\rho_1}=\dfrac{2\rho_1\rho_2}{\sqrt{\chi}} - \rho_1,&\\
\label{eq:inv_simple_2}
&\dot{\rho_2}=-\dfrac{\rho_1^2}{\sqrt{\chi}} + \dfrac{1}{4 \rho_2} + \dfrac{\zeta(t)}{\sqrt{2}},&
\end{eqnarray}
and these equations can be used to improve the above bounds in the limit $\chi \gg 1$. In particular, for the intensity of the pumped mode, we analytically obtain $\langle n_2 \rangle \geq 5 \chi/16$~\cite{SM}, which is closer to DNS results, see Fig.~\ref{fig:inverse_n}a.

The reduction in the number of degrees of freedom 
allows us to use larger values of parameter $d$ in the limit $\chi \gg 1$ since the size of the optimization space is also reduced. Moreover, we were able to obtain upper bounds ($d=32$) for correlation functions $\langle n_2 \rangle$ and $\langle n_1^2 \rangle$, which are close to the lower bounds, see Figs.~\ref{fig:inverse_n}a and \ref{fig:inverse_n}b. To estimate the higher moments of both modes, we will further use the values of lower bounds obtained for $d=32$. In this way, for the pumped mode, we found the usual scaling $\langle n_2^k\rangle \propto \langle n_2 \rangle^k$~\cite{SM}. In the intervals between bursts, the amplitude $\rho_1$ is close to zero, and, according to Eq.~(\ref{eq:inverse_b2}), one can expect that the statistics of the pumped mode is close to Gaussian. The value $\langle n_2 \rangle$ can be estimated as a diffusion displacement during the time between bursts, $\langle n_2 \rangle \sim \chi$, in agreement with the results reported earlier. The DNS data for the probability density function confirm this qualitative analysis, although the agreement is not perfect, see Fig.~\ref{fig:4}b.

For the dissipating mode, we found that $\langle n_1^k\rangle \propto \chi^{k-1}$~\cite{SM}, and such scaling is a fingerprint of intermittence discussed above. This suggests that in the limit $\chi \gg 1$, the distribution function of the dissipating mode has the following form:
\begin{equation}\label{eq:p1}
    P(n_1/\chi) = \dfrac{1}{\chi} F(n_1/\chi).
\end{equation}
DNS confirms this hypothesis and Fig.~\ref{fig:4}c shows the reduced probability density function $F(x)$, which is proportional to $1/x$ for $x\ll1$ and has an exponential cutoff for $x\gtrsim1$.

A simplified model for the dynamics of the dissipating mode explains the exponent $-1$ of the power law and resolves the singularity at $x\to0$. During the burst, the amplitude $\rho_2$ of the pumped mode quickly diminishes (see Fig. \ref{fig:4}a), therefore the second term  in Eq. (\ref{eq:inverse_b1}) dominates and the subsequent dynamics of the dissipating mode corresponds to the exponential decay, $\dot{x}=-2x$. For a given $x$, the probability density is determined by the time spent in the vicinity of $x$ and thus proportional to $F(x)\propto1/|\dot{x}|\propto1/x$.
The exponential dynamics is valid until the amplitude of the pumped mode is below $\sim\sqrt{\chi}$.  In this regime, the pumped mode grows due to diffusion and reaches the level of $\sqrt{\chi}$ in time of the order of $\chi$, so the dependence $F(x) \propto 1/x$ holds for $x \gtrsim e^{-2\chi}$. This value should be taken as the lower limit of integration in the expression for the total probability to resolve the singularity. Since the minimum value of $x$ depends on the parameter $\chi$, the height of the first bin on the histograms is proportional to $\chi$. Except for this, the shape of the curve $F(n_1/\chi)$ is universal, see Fig.~\ref{fig:4}c.

When calculating the positive moments of $x$, the singularity disappears and the lower limit of integration can be replaced by zero.  From the condition $\langle n_1 \rangle = 1$ we obtain $\int_0^{\infty} x F(x) dx = 1$, and so the function has the form $F(x) \simeq a x^{-1} e^{-a x}$. Numerical fitting leads to $a \approx 1.7$, and the corresponding curve is shown in Fig.~\ref{fig:4}c by a dashed line.

\section{Conclusion}
\label{sec:conclusion}

So what have we learned about the statistics of the far from the equilibrium state of the system using SoS programming? We obtained computer-aided analytical and numerical bounds for correlation functions in the system of two interacting modes in the regimes corresponding to both direct and inverse energy cascades. The bounds revealed the scaling of mode intensities and their higher moments with the Reynolds number $\chi$, which shows how far the system is from the energy equipartition corresponding to the thermal equilibrium. By combining scaling dependence with DNS, we collapsed the probability densities of dissipating and pumped modes on the universal curves in a highly intermittent regime of the inverse energy cascade and determined their shapes. Analyzing cross-mode correlations, we showed that the relative phase between modes tends to $\pi/2$ and $-\pi/2$ in the direct and inverse energy cascades, respectively, and estimated the rates of these transitions.


The method  applied does not work well in the opposite limit of low Reynolds number $\chi \ll 1$. This can be explained by the fact that the term produced by the white noise forcing in the optimization problems (\ref{eq:V_L}) has a relatively small amplitude and, therefore, the expression that is analyzed practically coincides with the case of a deterministic problem, which has a trivial solution $b_1=b_2=0$, and for this reason the lower bound does not exist. Methods for overcoming this issue are discussed in Ref.~\cite{fantuzzi2016bounds}, but we did not use them since the regime $\chi \ll 1$ is less interesting for us from the physical point of view.

Let us also note that the direct and inverse cascades are very different, although the equations at first glance may seem similar. In the direct cascade, the dissipating mode is excited by the pumped mode in an additive way. For the inverse cascade, this process is multiplicative, and the general experience suggests that this regime is more complicated because of higher temporal intermittency. Our analysis supports this intuition: in the inverse cascade we were able to obtain only lower bounds for the correlation functions, while for the direct cascade, we obtained both lower and upper bounds, and they are so close that we can determine the numerical values of the correlation functions with an accuracy comparable to DNS.

It would be promising to extend the present study to other shell models of turbulence with longer chains of resonantly interacting modes~\cite{Falkovich2021Fibonacci, shavit2022emerging}. The main difficulty is that for the long chains the extreme modes in the correlation function of interest will be coupled to the modes adjacent to them, and therefore the dynamic equations for the subset of modes involved in the correlation function are not closed. A similar problem arose when applying the sum-of-squares method to the Galerkin expansion of the Navier-Stokes equation in Ref.~\cite{chernyshenko2014polynomial}.  Although the generalization is not straightforward, the development of a complementary approach for this analysis significantly advances this area of research.


\acknowledgments
The work was supported by the Excellence Center at WIS, grant  662962 of the Simons  Foundation, grant 075-15-2022-1099 of the Russian Ministry of Science and Higher Educations, grant 823937 and 873028 of the EU Horizon 2020 programme, the BSF grants 2018033 and 2020765. V.P. is grateful to the Weizmann Institute, where most of the work was done, for their hospitality. E.M. thanks Benoziyo Endowment Fund for the Advancement of Science for funding his visit to the Weizmann Institute of Science. DNS were performed on the cluster of the Landau Institute.

%

\end{document}


\title{Supplemental Material: Sum-of-squares bounds on correlation functions\\in a minimal model of turbulence}
\date\today

\author{Vladimir Parfenyev$^{1,2}$, Evgeny Mogilevskiy$^{3,4}$, and Gregory Falkovich$^{1,4}$}

\affiliation{$^{1}$Landau Institute for Theoretical Physics, 142432 Chernogolovka, Russia}
\affiliation{$^{2}$National Research University Higher School of Economics, Faculty of Physics, 101000 Moscow, Russia}
\affiliation{$^{3}$Lomonosov Moscow State University, Faculty of Mechanics and Mathematics, 11992 Moscow, Russia}
\affiliation{$^{4}$Weizmann Institute of Science, 76100 Rehovot, Israel}

\begin{abstract}
Here we discuss technical details on analytical and numerical computations of bounds for the correlation functions in the model of two resonantly interacting oscillators defined in the main text.
\end{abstract}

\maketitle

\section*{General relations}
Let us remind that the direct and inverse energy cascades for the considered two-mode system are governed by the following systems of equations
\begin{eqnarray}
\label{eq:sm_dir}
&\displaystyle \dot{b}_1=-\frac{2ib_1^*b_2}{\sqrt{\chi}} +\xi(t), \quad
\displaystyle \dot{b}_2=-\frac{ib_1^2}{{\sqrt{\chi}}} -b_2,&\\
\label{eq:sm_inv}
&\displaystyle \dot{b}_1=-\frac{2ib_1^*b_2}{\sqrt{\chi}} -b_1, \quad
\displaystyle \dot{b}_2=-\frac{ib_1^2}{{\sqrt{\chi}}} +\xi(t),&
\end{eqnarray}
where $\xi(t)$ is a complex Gaussian random force with zero mean $\langle \xi(t) \rangle = 0$ and the variance $\langle \xi(t) \xi^*(t') \rangle = \delta(t-t')$, and the asterisk means complex conjugate. To use the relations from Sec. II of the main text, we rewrite the equations in terms of real four-dimensional vector $\bm{x} = (x_1, x_2, x_3, x_4)^T$, where $b_1=x_1+ix_2$, $b_2=x_3+ix_4$. Then, for the direct cascade, we obtain
\begin{equation}
\dot{x_i} = f_i (\bm x) + \sigma_{ij}(\bm x) \xi_j (t),\qquad
\bm{f}=\left(
\begin{array}{c}
-\dfrac{2}{\sqrt{\chi}}(x_2x_3-x_1x_4)\\
-\dfrac{2}{\sqrt{\chi}}(x_1x_3+x_2x_4)\\
\dfrac{2}{\sqrt{\chi}}x_1x_2-x_3\\
-\dfrac{1}{\sqrt{\chi}}(x_1^2-x_2^2)-x_3
\end{array}
\right),\quad
(\sigma_{ij})=\left(
\begin{array}{cc}
\dfrac{1}{\sqrt{2}}&0\\
0&\dfrac{1}{\sqrt{2}}\\
0&0\\
0&0
\end{array}
\right),\quad
\bm{\xi}=
\left(
\begin{array}{c}
\xi_1\\
\xi_2
\end{array}
\right),
\end{equation}
where $\xi_i(t)$ is a real Gaussian noise with zero mean $\langle \xi_i(t) \rangle = 0$ and the variance $\langle \xi_i (t) \xi_j (t') \rangle =\delta_{ij} \delta(t-t')$, and we sum over repeated indices. Similarly, for the inverse cascade, we find
\begin{equation}
\dot{x_i} = f_i (\bm x) + \sigma_{ij}(\bm x) \xi_j (t),\qquad
\bm{f}=\left(
\begin{array}{c}
-\dfrac{2}{\sqrt{\chi}}(x_2x_3-x_1x_4)-x_1\\
-\dfrac{2}{\sqrt{\chi}}(x_1x_3+x_2x_4)-x_2\\
\dfrac{2}{\sqrt{\chi}}x_1x_2\\
-\dfrac{1}{\sqrt{\chi}}(x_1^2-x_2^2)
\end{array}
\right),\quad
(\sigma_{ij})=\left(
\begin{array}{cc}
0&0\\
0&0\\
\dfrac{1}{\sqrt{2}}&0\\
0&\dfrac{1}{\sqrt{2}}
\end{array}
\right),\quad
\bm{\xi}=
\left(
\begin{array}{c}
\xi_1\\
\xi_2
\end{array}
\right).
\end{equation}
Next, the terms with an auxiliary function $Q(\bm{x})$ in the optimization problem (4) for the direct and inverse cascades are
\begin{equation}
\dfrac{1}{4}\left[\partial_{11}Q+\partial_{22}Q\right]+f_i\partial_iQ, \qquad \dfrac{1}{4}\left[\partial_{33}Q+\partial_{44}Q\right]+f_i\partial_iQ,
\end{equation}
with the corresponding vectors $\bm f$. They are the same for the optimization problem (5) but have the opposite sign. If $Q(\bm{x})$ is polynomial, the optimization problem's expression is also polynomial.

One can also rewrite the dynamical equations in terms of real variables
\begin{equation}\label{eq:sm_transformation}
\rho_1=|b_1|,\quad \rho_2=|b_2|,\quad \theta=\arg(b_1^2b_2^*).
\end{equation}
The right-hand sides of the corresponding equations are no longer polynomial (see the main text), and some preparatory steps are needed before posing the optimization problem. As an example, let us consider the direct cascade. First, we introduce $s(t)=\sin[\theta(t)]$ and $c(t)=\cos[\theta(t)]$, and then the dynamic equations for the vector $\bm x=(\rho_1,\ \rho_2,\ s)^T$ are
\begin{equation}\label{eq:sm_dir_r}
\dot{x_i} = f_i (\bm x) + \sigma_{ij}(\bm x) \zeta_j (t),\qquad
\bm{f}=\left(
\begin{array}{c}
-\dfrac{2\rho_1\rho_2 s}{\sqrt{\chi}}+\dfrac{1}{4\rho_1}\\
\dfrac{2\rho_1^2 s}{\sqrt{\chi}}-\rho_2\\
\dfrac{\rho_1^2-4\rho_2^2}{\rho_2\sqrt{\chi}}(1-s^2)
\end{array}
\right),\quad
(\sigma_{ij})=\left(
\begin{array}{cc}
\dfrac{1}{\sqrt{2}}&0\\
0&0\\
0&\dfrac{\sqrt{2}c}{\rho_1}
\end{array}
\right),\quad
\bm{\zeta}=
\left(
\begin{array}{c}
\zeta_1\\
\zeta_2
\end{array}
\right).
\end{equation}
Next, we consider an auxiliary function $Q$, which is polynomial with respect to $(\rho_1,\ \rho_2,\ s)$. Then, the terms involving $Q$ in the optimization problem (4) are
\begin{equation}
\frac{1}{2} \sigma_{kj} \partial_k (\sigma_{ij} \partial_i Q) + f_i \partial_i Q =\dfrac{1}{4}\partial_{\rho_1\rho_1}Q+\dfrac{1}{\rho_1^2}\left[(1-s^2)\partial_{ss}Q-s\partial_sQ\right]+ f_i \partial_i Q.
\end{equation}
The obtained expression is not polynomial, as the terms with $\partial_{ss}Q$ have the denominator of $\rho_1^2$ and the term $f_3\partial_sQ$ has the denominator of $\rho_2$. However, these denominators are positive; thus, the non-negativity of the expression of interests in the left-hand side of relation (4) is equivalent to the non-negativity of this expression multiplied by $\rho_1^2\rho_2$, which is polynomial. Thus, the optimization problem for a lower bound in the direct cascade is
\begin{equation}
 \max_{ Q(\rho_1,\rho_2,s)} L: \; \rho_1^2\rho_2\left[\dfrac{1}{4}\partial_{\rho_1\rho_1}Q+\dfrac{1}{\rho_1^2}\left[(1-s^2)\partial_{ss}Q-s\partial_sQ\right]+ f_i \partial_i Q+ \phi - L\right]\in SoS,
\end{equation}
with $\bm{f}$ from (\ref{eq:sm_dir_r}). Similarly, for the inverse cascade, one obtains:
\begin{equation}
\begin{array}{c}
\displaystyle \max_{ Q(\rho_1,\rho_2,s)} L: \; \rho_2^2\left[\dfrac{1}{4}\partial_{\rho_1\rho_1}Q+\dfrac{1}{4\rho_2^2}\left[(1-s^2)\partial_{ss}Q-s\partial_sQ\right)+ f_i \partial_i Q+ \phi - L\right]\in SoS,\\
 \bm{f}=\left(-\dfrac{2\rho_1\rho_2 s}{\sqrt{\chi}}-\rho_1,\
\dfrac{2\rho_1^2 s}{\sqrt{\chi}}+\dfrac{1}{4\rho_2},\
\dfrac{\rho_1^2-4\rho_2^2}{\rho_2\sqrt{\chi}}(1-s^2)
 \right).
 \end{array}
\end{equation}
One calculates the upper bounds in the same way.

Finally, one can improve the bounds by considering the constraint $1-s^2\geq 0 $. We replace the condition that the expression in the left-hand side of the optimization problem is non-negative by that it is larger than non-negative polynomial $(1-s^2)$ multiplied by another non-negative (SoS) polynomial $S(\rho_1,\rho_2,s)$. Ultimately, the optimization problem for the lower bound in the direct cascade is
\begin{equation}
 \max_{S(\rho_1,\rho_2,s)\in SoS,\, Q(\rho_1,\rho_2,s)} L: \; \rho_1^2\rho_2\left[\dfrac{1}{4}\partial_{\rho_1\rho_1}Q+\dfrac{1}{\rho_1^2}\left[(1-s^2)\partial_{ss}Q-s\partial_sQ\right]+ f_i \partial_i Q+ \phi - L\right]-S(\rho_1,\rho_2,s)(1-s^2)\in SoS.
\end{equation}
Straightforward transformations give the problems for the inverse cascade and the upper bound.


\section*{Transformation to real variables}

Here we discuss how to rewrite equations (\ref{eq:sm_inv}) for the inverse cascade in term of real variables (\ref{eq:sm_transformation}). The reasoning for the direct cascade is similar. First, let us introduce $b_1 = \rho_1 e^{i \phi_1}$, $b_2 = \rho_2 e^{i \phi_2}$, and then $\theta = 2 \phi_1 - \phi_2$. The complex noise $\xi(t) = (\xi_1(t) + i \xi_2(t))/\sqrt{2}$, where the real and imaginary parts are uncorrelated and $\langle \xi_i (t) \xi_j (t') \rangle = \delta_{ij} \delta(t-t')$. After a trivial substitution into the equations (\ref{eq:sm_inv}), we find
\begin{eqnarray}
&\dot{\rho_1} = -\dfrac{2 \rho_1 \rho_2 \sin \theta}{\sqrt{\chi}} - \rho_1, \qquad \dot{\rho_2} = \dfrac{\rho_1^2 \sin \theta}{\sqrt{\chi}} + \dfrac{1}{\sqrt{2}} [\xi_1 \cos \phi_2 + \xi_2 \sin \phi_2],&\\
&\dot{\phi_1} = - \dfrac{2 \rho_2 \cos \theta}{\sqrt{\chi}}, \qquad \dot{\phi_2} = - \dfrac{\rho_1^2 \cos \theta}{\rho_2 \sqrt{\chi}} + \dfrac{1}{\rho_2 \sqrt{2}} [-\xi_1 \sin \phi_2 + \xi_2 \cos \phi_2],&
\end{eqnarray}
and the multiplicative noise must be interpreted in Stratonovich's sense. Next, we move to Ito calculus, which will produce an additional drift term, but the noise $\xi_{1,2}$ and the phase $\phi_2$ will become uncorrelated. Now, we can interpret the terms $\xi_1 \cos \phi_2$ and $\xi_2 \sin \phi_2$ as independent random variables $X_1 \sim \mathcal{N}(0, \cos^2 \phi_2)$ and $X_2 \sim \mathcal{N}(0, \sin^2 \phi_2)$, where $\mathcal{N}(\mu, \sigma^2)$ stands for the normal distribution, and their sum $X_1 + X_2$ can be considered as a new random variable $\zeta_1 \sim \mathcal{N}(0,1)$. In the same way, we can define $\zeta_2 = -\xi_1 \sin \phi_2 + \xi_2 \cos \phi_2 \sim \mathcal{N}(0,1)$, and it does not correlate with $\zeta_1$, i.e. $\langle \zeta_1 \zeta_2 \rangle = 0$. As a result, we arrive at a closed system of equations
\begin{equation}
\dot{\rho_1}=-\dfrac{2\rho_1\rho_2\sin\theta}{\sqrt{\chi}} - \rho_1, \quad
\dot{\rho_2}=\dfrac{\rho_1^2 \sin\theta}{\sqrt{\chi}} + \dfrac{1}{4 \rho_2} + \dfrac{\zeta_1(t)}{\sqrt{2}}, \quad
\dot{\theta}=\dfrac{\rho_1^2-4\rho_2^2}{\rho_2 \sqrt{\chi}}\cos\theta+\dfrac{\zeta_2(t)}{\sqrt{2} \rho_2},
\end{equation}
and note that the overall phase drops out.

\section*{Analytical bounds for correlation functions}
This section discusses how SoS optimization allows us to impose bounds for the correlation functions analytically. We introduce mode intensities $n_1 = |b_1|^2$, $n_2 = |b_2|^2$ and the energy flux $J = i (b_1^2 b_2^* - b_1^{*2}b_2)/\sqrt{\chi}$.   We will explain in detail how to obtain the upper bound for $\langle n_1\rangle$ in direct cascade and briefly describe our steps for other cases.

\subsection*{Direct Cascade}
To prove the upper bound $\langle n_1 \rangle \leq U$, we consider the auxiliary function $Q = A n_1 + B n_2 + C J + D n_1^2 + E n_2^2 + F n_1 n_2$, where $A$-$F$ are some constants. This form is a general form of a polynomial in $n_1$, $n_2$, $J$ with the degree in mode amplitudes less than or equal to $d=4$. As explained in the main text, to find the upper bound, we should solve the optimization problem (5), which reads
\begin{equation}\label{eq:dir_n1_U}
\begin{aligned}
    \min_{A-F} U: U - n_1 - A (2J + 1) + B(J + 2n_2) - C \left( \dfrac{8 n_1 n_2}{\chi} - \dfrac{2 n_1^2}{\chi} - J \right)
    - 4 D (n_1 J + n_1) +\\ + 2E (n_2 J + 2 n_2^2) - F (-J n_1 -2 n_1 n_2 + 2J n_2 +n_2) \in SoS.
\end{aligned}
\end{equation}
An alternative way to get this expression is to calculate $\langle dQ/dt \rangle$ and consider inequality $\langle U - n_1 - dQ/dt \rangle \geq 0$. In the statistical steady-state $\langle dQ/dt \rangle = 0$, and therefore this inequality implies the bound relation. Evaluating the expression for derivatives mean values using the Furutsu-Novikov formula
\begin{eqnarray*}
&\left\langle \dfrac{dn_1}{dt} \right\rangle = \langle 2J + 1 \rangle, \quad \left\langle \dfrac{dn_2}{dt} \right\rangle = \langle -J - 2n_2 \rangle, \quad  \left\langle \dfrac{dJ}{dt} \right\rangle = \left\langle \dfrac{8 n_1 n_2}{\chi} - \dfrac{2 n_1^2}{\chi} - J \right\rangle,&\\
&\left\langle \dfrac{dn_1^2}{dt} \right\rangle = \langle 4 n_1 J + 4 n_1 \rangle, \quad \left\langle \dfrac{dn_2^2}{dt} \right\rangle = \langle -2 n_2 J -4 n_2^2 \rangle, \quad \left\langle \dfrac{d n_1 n_2}{dt} \right\rangle = \langle -J n_1 -2 n_1 n_2 + 2J n_2 +n_2 \rangle.&
\end{eqnarray*}
we arrive to Eq.~(\ref{eq:dir_n1_U}). Then, the bound $U$ is the minimum number that satisfies $\langle U- n_1 - dQ/dt\rangle \geq 0$, and we replace checking the non-negativity of the polynomial by a stricter condition that the polynomial of interest is a sum of squares (SoS) of other polynomials. The idea of analytical solution is to provide an appropriate form of SoS and choose the values of coefficients $A$~--- $F$.

Next, we require that the terms containing $J$ be set to zero in the optimization problem (\ref{eq:dir_n1_U}) since the resulting expression must be SoS. We find $F=E=4D$ and $2A = B+C$, and therefore the optimization problem becomes
\begin{equation}
\begin{aligned}
    \min_{B,C,F} U: \dfrac{2C}{\chi} n_1^2 + 4F n_2^2 + 2 n_1 n_2 \left(F- \dfrac{4C}{\chi} \right) - n_1 (1+F) - n_2 (F-2B) + U - \dfrac{B+C}{2}  \in SoS.
\end{aligned}
\end{equation}
The solution to this problem can be found by the computer algorithm, and we can use this result to infer the form of the SoS polynomial. It turns out that the value of $U$ is minimal if SoS has the form $(\alpha n_1 + \beta n_2 + \gamma)^2$, where $\alpha<0$, $\beta>0$, $\gamma>0$ are some constants, and therefore
\begin{eqnarray*}
    &\alpha^2 = 2C/\chi, \quad \beta^2 = 4 F, \quad \alpha \beta = F - 4C/\chi,&\\
    &2 \alpha \gamma = -1-F, \quad 2 \beta \gamma = 2B -F, \quad \gamma^2 = U - \dfrac{B+C}{2}.&
\end{eqnarray*}
After straightforward calculations, we obtain
\begin{equation*}
   \dfrac{C}{\chi} = \dfrac{(2+\sqrt{3}) F}{4}, \quad \gamma = \dfrac{1+F}{\sqrt{4+2\sqrt{3}} \sqrt{F}}, \quad B = \dfrac{F}{2} + (\sqrt{3}-1)(F+1),
\end{equation*}
and then
\begin{equation}
    U = \min_{F \geq 0} \dfrac{8 - 4\sqrt{3} + 4 F(3-\sqrt{3}) + F^2 (6+ (2+\sqrt{3})\chi)}{8 F} = \dfrac{1}{2} \left(3 -\sqrt{3} + \sqrt{12 - 6 \sqrt{3}+\chi} \right).
\end{equation}
This result is given in the main text.

Similarly, one can find the upper bound for $\langle n_1^2 \rangle$ and $\langle n_2^2 \rangle$. In these cases, we assume that the SoS polynomials have the same form $(\alpha n_1 + \beta n_2 + \gamma)^2$. The solution to the first problem is given in the main text, and for the second problem, one finds
\begin{equation}\label{eq:sm_dir_n22}
    \langle n_2^2 \rangle \leq \dfrac{12 + \chi + \sqrt{36 + 24 \chi + \chi^2}}{96}.
\end{equation}
In the limit $\chi \gg 1$, this bound significantly overestimates the correlation function (gives wrong scaling with $\chi$). Still, it can be improved numerically by taking into account the higher order terms with respect to the mode amplitudes in the auxiliary function $Q$, see Fig.~1b in the main text. In the limit $\chi \ll 1$, Eq. (\ref{eq:sm_dir_n22}) gives $\langle n_2^2 \rangle \leq 3/16$, and the scaling in this limit is consistent with DNS.

The methodology for obtaining lower bounds for correlation functions remains unchanged, except for the change in the inequality sign, see expression (4) in the main text. For the dissipating mode, we find $\langle n_2^2 \rangle \geq 1/16$ by considering the SoS polynomial of the form $(n_2 + \gamma)^2$. Note that the same inequality directly follows from the positive variance $\langle n_2^2\rangle \geq \langle n_2 \rangle^2 = 1/16$. One needs to consider more complex SoS polynomial expressions to get reliable bounds for the pumped mode. In particular, to obtain the lower bound $\langle n_1 \rangle \geq L$ we consider SoS polynomial, which has the form
\begin{equation*}
[\alpha (x_3^2+x_4^2)-\beta]^2 + [-\gamma(x_1 x_3 + x_2 x_4)+\delta x_2]^2 + [\gamma (x_1 x_4 - x_2 x_3) + \delta x_1]^2,
\end{equation*}
where $b_1 = x_1 + i x_2$, $b_2 = x_3 + i x_4$. This SoS polynomial should be equal to the expression under optimization:
\begin{equation}
\begin{aligned}
\max_{A-F} L: n_1 - L + A(2J + 1) + B(-J - 2n_2) + C \left( \dfrac{8 n_1 n_2}{\chi} - \dfrac{2 n_1^2}{\chi} - J \right)
+ D(4 n_1 J + 4 n_1) + \\+ E(-2 n_2 J -4 n_2^2) + F (-J n_1 -2 n_1 n_2 + 2J n_2 +n_2) \in SoS.
\end{aligned}
\end{equation}
Thus, we readily obtain $C=0$, $F=E=4D$, and then the optimization problem can be simplified
\begin{equation}
\begin{aligned}
\max_{A,B,F} L: n_1 - L + A(2J + 1) + B(-J - 2n_2) + F n_1 -4 F n_2^2 + F (-2 n_1 n_2 +n_2) \in SoS.
\end{aligned}
\end{equation}
Next, by equating the coefficients for the same terms, we find
\begin{eqnarray*}
&\alpha^2 = -4F, \quad -2 \alpha \beta = -2B + F, \quad \beta^2 = -L + A, \quad \gamma^2 = -2 F, \quad \gamma \delta \chi^{1/2} = (2A - B), \quad \delta^2 = 1+F,&
\end{eqnarray*}
and by introducing $F=-w^2$, where $0 \leq w \leq 1$, one finally gets
\begin{equation}
    L = \max_{A,w} \left[ A - \dfrac{(4A + w^2 - 2 w \sqrt{2\chi (1-w^2)})^2}{16 w^2} \right] = \dfrac{\chi^{1/2}}{2\sqrt{2}}.
\end{equation}
To obtain the lower bound for $\langle n_1^2 \rangle$, we consider the SoS polynomial of the form
$[\alpha(x_1 x_3 + x_2 x_4) - \beta x_2]^2 + [-\alpha(x_1 x_4 - x_2 x_3) - \beta x_1]^2 + [\delta-\gamma (x_1^2 + x_2^2) ]^2$,
and a similar analysis leads to the answer given in the main text.

\subsection*{Inverse Cascade}

We found that we can get better bounds for correlation functions in the inverse cascade by considering dynamic equations in terms of real variables $\rho_1=|b_1|$, $\rho_2=|b_2|$, $\theta=\arg(b_1^2b_2^*)$
\begin{equation}
\dot{\rho_1}=-\dfrac{2\rho_1\rho_2\sin\theta}{\sqrt{\chi}} - \rho_1, \quad
\dot{\rho_2}=\dfrac{\rho_1^2 \sin\theta}{\sqrt{\chi}} + \dfrac{1}{4 \rho_2} + \dfrac{\zeta_1(t)}{\sqrt{2}}, \quad
\dot{\theta}=\dfrac{\rho_1^2-4\rho_2^2}{\rho_2 \sqrt{\chi}}\cos\theta+\dfrac{\zeta_2(t)}{\sqrt{2} \rho_2}.
\end{equation}
To prove the lower bound $\langle n_2 \rangle \equiv \langle \rho_2^2 \rangle \geq L$, we consider the auxiliary function $Q = \rho_2^2/4 + \rho_1^2/8 + \chi^{1/2} (\rho_2 \sin \theta)/4 - A \ln \rho_1$, where $A$ is some constant. This form was chosen by analyzing the results obtained by a computer algorithm. In the statistical steady-state $\langle dQ/dt \rangle = 0$, and therefore it is sufficient to show that $\langle \rho_2^2 - L + dQ/dt \rangle \geq 0$. The idea is to choose the constant $A$ so that the expression on the left-hand side is the sum of squares (SoS) of other polynomials. Next, we calculate
\begin{eqnarray*}
&\left\langle \dfrac{d \rho_2^2}{dt} \right\rangle = \left\langle \dfrac{2 \rho_1^2 \rho_2 \sin \theta}{\sqrt{\chi}} + 1 \right\rangle, \quad \left\langle \dfrac{d \rho_1^2}{dt} \right\rangle = \left\langle -\dfrac{4 \rho_1^2 \rho_2 \sin \theta}{\sqrt{\chi}} -2 \rho_1^2 \right\rangle,&\\
&\left\langle \dfrac{d }{dt} \rho_2 \sin \theta \right\rangle = \left\langle \dfrac{\rho_1^2 \sin^2 \theta}{\sqrt{\chi}} + \dfrac{\rho_1^2 - 4 \rho_2^2}{\sqrt{\chi}} (1-\sin^2 \theta)\right\rangle, \quad \left\langle \dfrac{d \ln \rho_1}{dt} \right\rangle = \left\langle -\dfrac{2 \rho_2 \sin \theta}{\sqrt{\chi}} - 1 \right\rangle,&
\end{eqnarray*}
and then the expression $\langle\rho_2^2 -L +  dQ/dt \rangle = \langle(\rho_2 \sin \theta + A/\sqrt{\chi})^2 - A^2/\chi -L +1/4 + A\rangle$. It is positive when the term outside the parentheses is zero, and then we find the maximum value for the constant $L$:
\begin{equation}
    L = \max_A \left( A + \dfrac{1}{4} - \dfrac{A^2}{\chi} \right) = \dfrac{\chi+1}{4}.
\end{equation}

The lower bound for $\langle n_2^2 \rangle$ can be obtained from the condition of positive variance, $\langle n_2^2 \rangle \geq \langle n_2 \rangle^2 \geq (\chi+1)^2/16$. A similar argument for the first mode leads to a result that can be significantly improved. To do this, we again consider the auxiliary function $Q= - 4\chi (\ln \rho_1) /3+ \rho_1^4/6 + 2 \rho_1^2 \rho_2^2/3 - \chi^{1/2} \rho_1^2 \rho_2 \sin \theta/3 + 2 \rho_2^4/3 + (4 - \chi) \rho_2^2/3 + 2 \rho_1^2/3 + 2 \chi^{1/2} \rho_2 \sin \theta/3$, and using
\begin{eqnarray*}
    &\left\langle \dfrac{d \rho_1^4}{dt} \right\rangle = \left\langle 4 \rho_1^4 \left(- \dfrac{2 \rho_2 \sin \theta}{\sqrt{\chi}} - 1 \right) \right\rangle, \quad \left\langle \dfrac{d \rho_2^4}{dt} \right\rangle = \left\langle \dfrac{4 \rho_1^2 \rho_2^3 \sin \theta}{\sqrt{\chi}} + 4 \rho_2^2 \right\rangle,&\\
    &\left\langle \dfrac{d }{dt} \rho_1^2 \rho_2 \sin \theta \right\rangle = \left\langle 2 \rho_1^2 \rho_2 \sin \theta \left(- \dfrac{2 \rho_2 \sin \theta}{\sqrt{\chi}} - 1 \right) + \dfrac{\rho_1^4 \sin^2 \theta}{\sqrt{\chi}} + \dfrac{\rho_1^2 (\rho_1^2 - 4 \rho_2^2)}{\sqrt{\chi}} (1 - \sin^2 \theta) \right\rangle,&\\
    &\left\langle \dfrac{d }{dt} \rho_1^2 \rho_2^2 \right\rangle = \left\langle 2 \rho_1^2 \rho_2^2 \left(- \dfrac{2 \rho_2 \sin \theta}{\sqrt{\chi}} - 1 \right) + 2 \rho_1^2 \left( \dfrac{\rho_1^2 \rho_2 \sin \theta}{\sqrt{\chi}} + \dfrac{1}{2} \right) \right\rangle,&
\end{eqnarray*}
we find that
\begin{equation}
    \left\langle \rho_1^4 - L + \dfrac{dQ}{dt} \right\rangle = \left\langle \dfrac{8}{3} \left( \rho_2 \sin \theta + \dfrac{\sqrt{\chi}}{2} \right)^2 + \dfrac{\chi+4}{3} - L\right\rangle.
\end{equation}
Thus, if $L = (\chi+4)/3$, then $\left\langle \rho_1^4 - L + dQ/dt \right\rangle \geq 0$, and using $\langle dQ/dt \rangle = 0$, we finally get $\langle n_1^2 \rangle \geq (\chi+4)/3$.

In the limit $\chi \gg 1$, we can simplify the dynamic equations by assuming that the relative phase between modes is locked on $\theta = -\pi/2$ (see the main text for details). Then
\begin{equation}
\dot{\rho_1}=\dfrac{2\rho_1\rho_2}{\sqrt{\chi}} - \rho_1, \quad
\dot{\rho_2}=-\dfrac{\rho_1^2}{\sqrt{\chi}} + \dfrac{1}{4 \rho_2} + \dfrac{\zeta(t)}{\sqrt{2}},
\end{equation}
and we can improve the lower bound for the intensity of the pumped mode. To do this, we consider the polynomial combination $\rho_2 (\rho_2^2 -L + \langle dQ/dt \rangle)$, where the function $Q = A \rho_1^4 + B \rho_1^2 \rho_2^2 + C \rho_2^4 + D \rho_1^2 \rho_2 + E \rho_2^3 +F \rho_1^2 + G \rho_2^2 + H \ln \rho_1$ and $A-H$ are some constants. By adjusting the values of these constants, we represent expression under the averaging sign in $\langle\rho_2 (\rho_2^2 -L +  dQ/dt )\rangle$ as sum-of-squares (SoS) of other polynomials, i.e., we solve the optimization problem
\begin{equation}
\begin{aligned}
\max_{A-H} L: \langle \rho_2 (\rho_2^2 -L + dQ/dt  )\rangle \in SoS,
\end{aligned}
\end{equation}
where the form of SoS polynomial can be found in advance by using a computer algorithm. In our case, it is equal to $(-\alpha \rho_1 \rho_2^2 + \alpha \chi^{1/2} \rho_1 \rho_2 - \beta \rho_1)^2$, and after the straightforward but lengthy calculations we find the solution $L = 5 \chi/16$.

\section*{Details of numerical calculations}

The numerical algorithm is sensitive to the scaling of variables. We used several forms of Eqs. (\ref{eq:sm_dir}) and (\ref{eq:sm_inv})  obtained by rescaling of the variables. The general idea is to use a scaling in which the average mode intensities are of the order of unity. Otherwise, the examined numerical procedures may not find feasible solutions.

We use the auxiliary function $Q$ in the form of polynomial of $n_1$, $n_2$, $J$ and represent it as a polynomial with respect to $x_1,\ x_2,\ x_3,\ x_4$, where $b_1=x_1+ix_2$ and $b_2=x_3+ix_4$. We fix the degree $d$ of $Q$ as a polynomial of $x_i$; the value of $d$ is a key internal parameter for numerical optimization. An increase of the value of $d$ increases the number of the undetermined coefficients in the ansatz $Q$ and the dimension of the optimization space (number of monomials that form the summands in SoS). The numerical procedure fails if the latter is large enough. The maximal value of $d$ that leads to a feasible solution depends on $\chi$. We obtained a solution for the largest values of $d=16$ for $\chi\sim1$; the results in the diapason of interest $\chi\gg1$ are obtained for $d=10$. For direct cascade, we observe close values for upper and lower bounds in this case.


An effective way to decrease the dimension of the optimization space is to decrease the number of independent variables. The variable transformation from $b_1=x_1+ix_2,\ b_2=x_3+ix_4$ to $\rho_1,\ \rho_2,\ \theta$  (\ref{eq:sm_transformation}) drops out the overall phase and decrease the number of variables from 4 to 3. We use Eqs. (11)~--- (13) from the main text to impose bounds on $\theta$ only since other results can be obtained in the initial formulation with enough accuracy for $\chi>10^2$. When the variables $\rho_1,\ \rho_2,\ \theta$ are used, the auxillary function $Q$ is a polynomial with respect to $\rho_1, \rho_2, s$, $s=\sin\theta$, with the total degree of $d$. The function $S$ used for the constraint is an independent arbitrary polynomial; the numerical algorithm automatically adjusts its degree.

Further simplification of the optimization problem appears when we lock the phase at its asymptotic value $\theta=\pm\pi/2$. We use this approach to study inverse cascade at $\chi\gg1$.
Taking $Q$ as a polynomial of the degree $d$ with respect to $\rho_1$, $\rho_2$ in a linear combination with $\ln \rho_1$, we tested $d$ up to 32 and obtained the results presented in the main text.
This way, we obtain upper bounds for the inverse cascade in the asymptotic limit.

\section*{Higher moments in inverse cascade}
For the inverse cascade, SoS programming gives upper bounds for the moments of marginal distribution function only for the reduced system (21), (22) from the main text where the relative phase is locked at $-\pi/2$. For the dissipating mode, we obtained both upper and lower bounds for correlators $\langle n_1^k\rangle$, $k=1,\dots, 5$ for $\chi>10^2$. The lower bounds follow the scaling $\langle n_1^k\rangle\sim\chi^{k-1}$; the upper one slowly goes towards it (Table \ref{tab:inv_n1}).

\begin{table}[]
    \centering
    \begin{tabular}{|c|c||c||c|c||c|c||c|c||c|c||}
\hline
$\chi$ &	$\log_{10}\chi$ &	$\langle n_1\rangle $ &	$\langle n_1^2\rangle_L/\chi$ & $\langle n_1^2\rangle_U/\chi$ & $\langle n_1^3\rangle_L/\chi^2$ & $\langle n_1^3\rangle _U/\chi^2$ & $\langle n_1^4\rangle _L/\chi^3$ & $\langle n_1^4\rangle_U/\chi^3$ & $\langle n_1^5\rangle_L/\chi^4$ & $\langle n_1^5\rangle_U/\chi^4$ \\
\hline
158.49&	2.2&	1&    0.535& 	0.573&	0.374&	0.643&	0.301&	1.230& 	0.267&	2.843\\ \hline
398.11&     2.6&	1&	0.524&	0.557&	0.343&	0.706&	0.248&	0.953&	0.192&	1.622\\ \hline
1000&       3&   	1&  	0.519&	0.564&	0.331&      0.771&	0.229&	1.081&	0.166&	1.711\\ \hline
2511.89&	3.4&	1&	0.517&	0.551&	0.325&	0.616&	0.221&	0.920&      0.156&	1.196\\ \hline
6309.57&	3.8&	1&	0.516&	0.540&      0.324&	0.617&	0.217&	0.629&	0.153&	0.758\\ \hline
15848.93&	4.2&	1&	0.516&	0.542&	0.322&      0.601&	0.216&	0.619&	0.151&	0.862\\ \hline
39810.72&	4.6&	1&	0.516&	0.536& 	0.323&	0.444&      0.217&	0.523&	0.148&	0.979\\ \hline
100000&    	5&	1&	0.516&  	0.534&      0.327&      0.522&      0.217&      0.444&      0.150&      0.430\\
\hline
    \end{tabular}
    \caption{Bounds for high moments of the dissipating mode in inverse cascade for $\chi\gg1$ with the relative phased locked at $\theta=-\pi/2$}
    \label{tab:inv_n1}
\end{table}

For the pumped mode, we obtained lower bounds for the first five moments $\langle n_2^k\rangle$, $k=1,\dots, 5$ and the upper bound for $\langle n_2\rangle$ only, thus only lower bounds for the higher moments are available. These bounds scale as $\langle n_2\rangle^k$ while $\chi\to\infty$, the normalized lower bound $\langle n_2^k\rangle_L/(k!\langle n_2\rangle_U^k)$, where subindexes $L$ and $U$ mean ``lower'' and ``upper'', is below unity allowing the statistics to be Gaussian (Table \ref{tab:inv_n2}).

\begin{table}[]
    \centering
    \begin{tabular}{|c|c||c|c||c|c||c|c||c|c||c|c||}
\hline
$\chi$ &	$\log_{10}\chi$ &	$\langle n_2\rangle_L $ &$\langle n_2\rangle_U $ &	$\langle n_2^2\rangle_L$ & $\langle n_2^2\rangle_L/(2!\langle n_2\rangle_U^2)$ & $\langle n_2^3\rangle_L$ & $\langle n_2^3\rangle_L/(3!\langle n_2\rangle_U^3)$ & $\langle n_2^4\rangle _L$ & $\langle n_2^4\rangle_L/(4!\langle n_2\rangle_U^4)$ & $\langle n_2^5\rangle_L$ & $\langle n_2^5\rangle_L/(5!\langle n_2\rangle_U^5)$ \\
\hline
158.49&	2.2&	0.319&	0.335&	0.186&	0.826&	0.139&	0.615&	0.121&	0.397&	0.115&	0.225\\ \hline
398.11&	2.6&	0.318&	0.341&	0.181&	0.778&	0.132&	0.552&	0.109&	0.335&	0.100&	0.179\\ \hline
1000& 	3&	0.317&	0.343&	0.179&	0.763&	0.129&	0.532&	0.105&	0.317&	0.093&	0.164\\ \hline
2511.89&	3.4&	0.317&	0.341&	0.178&	0.766&	0.127&	0.534&	0.103&	0.319&	0.091&	0.164\\ \hline
6309.57&	3.8&	0.316&	0.342&	0.178&	0.760&	0.126&	0.525&	0.102&	0.311&	0.090&	0.160\\ \hline
15848.93&	4.2&	0.316&	--&	      0.177&	--&	      0.125&	--&	      0.102&	--&	      0.089&	--\\ \hline
39810.72&	4.6&	0.315&	--&	      0.177&	--&	      0.125&	--&	      0.101&	--&	      0.089&	--\\ \hline
100000&	5&	0.315&	--&	      0.176&	--&	      0.125&	--&	      0.100&	--&	      0.088&	--\\ \hline
    \end{tabular}
    \caption{Bounds for high moments of the pumped mode in inverse cascade for $\chi\gg1$ with the relative phased locked at $\theta=-\pi/2$. Dashes mean no results are available for the bound.}
    \label{tab:inv_n2}
\end{table}


\section*{Three modes system}
We consider a triplet as another simplified shell-model of turbulence. The governing equations are
\[
\begin{array}{c}
\dot{b_1}=-\dfrac{ib_2^*b_3}{\sqrt{\chi}}-b_1\\
\dot{b_2}=-\dfrac{ib_1^*b_3}{\sqrt{\chi}}-b_2\\
\dot{b_3}=-\dfrac{ib_1b_2}{\sqrt{\chi}}+\xi(t)\\
\end{array}
\]
with the same notation as before. The equations describe the inverse cascade with the energy transfer from the pumped mode 3 to dissipating modes 1 and 2. For simplicity, we assume equals damping coefficients for dissipating modes. The energy flux is $J=(ib_1b_2b_3^*+ c.c.)/\sqrt{\chi}$, in the statistically steady state the flux is
$\langle J\rangle=1$, the dissipating modes intensities are $\langle n_{1,2}\rangle\equiv\langle |b_{1,2}|^2\rangle=1/2$. Using SoS programming, we impose bounds for the pumped mode and the second moment of the dissipating one. The analytical result for the lower bound of $\langle n_3\rangle$ comes from the consideration of the auxiliary function $Q$ as the polynomial on $n_1$, $n_2$, $n_3$, $J$ with the total degree $d=4$ with respect to $b_i$. The monomials and mean values of their derivatives obtained with Furutsu-Novikov formula are
\[
\begin{array}{c}
\left\langle\frac{d n_1}{dt}\right\rangle=\left\langle J - 2n_1 \right\rangle,\quad \left\langle\frac{d n_2}{dt}\right\rangle=\left\langle J - 2n_2 \right\rangle,\quad \left\langle\frac{d n_3}{dt}\right\rangle=\left\langle 1- J\right\rangle,\\
\left\langle\frac{d n_1^2}{dt}\right\rangle=\left\langle 2Jn_1 - 4n_1^2 \right\rangle,\quad \left\langle\frac{d n_2}{dt}\right\rangle=\left\langle 2Jn_2 - 4n_2^2 \right\rangle,\quad \left\langle\frac{d n_3}{dt}\right\rangle=\left\langle 4n_3- 2Jn_3\right\rangle,\\
\left\langle\frac{d n_1n_2}{dt}\right\rangle=\left\langle Jn_1+Jn_2 - 4n_1n_2 \right\rangle,\quad \left\langle\frac{d n_2n_3}{dt}\right\rangle=\left\langle Jn_3-Jn_2 - 2n_2n_3 +n_2\right\rangle,\quad \left\langle\frac{d n_3 n_1}{dt}\right\rangle=\left\langle Jn_3-Jn_1 - 2n_1n_3 +n_1\right\rangle,\\
\left\langle\frac{d J}{dt}\right\rangle=\left\langle \dfrac{2n_1n_3}{\chi}+\dfrac{2n_2n_3}{\chi} -\dfrac{2n_1n_1}{\chi}-J\right\rangle.
\end{array}
\]
Stating SoS in the form of $\alpha(n_1+n_2+\beta)^2+\gamma(n_1-n_2+\delta)^2$ gives
\[
\langle n_3\rangle \geq \frac{1+2\chi}{8}.
\]
For the second moment of dissipating modes $\langle n_{1,2}^2\rangle$ we found that the lower bounds increase with $\chi$ at $\chi\gg1$, but the calculations with $d\leq14$ did not show convergence to a distinct scaling.